\documentclass[12pt]{article}
\usepackage{txfonts}
\usepackage{natbib}
\usepackage{graphicx}
\usepackage{times}
\usepackage{wasysym}
\usepackage{color}
\usepackage{endnotes}
\bibpunct{(}{)}{;}{a}{}{,}
\date{August 13, 2015}

\newcommand{\Qn}{\mathcal Q^{\rm N}_{\odot}}
\newcommand{\Qnom}{\hbox{$\mathcal{Q}^{\rm N}_{\odot}$}}
\newcommand{\Snom}{\hbox{$\mathcal{S}^{\rm N}_{\odot}$}}
\newcommand{\Rnom}{\hbox{$\mathcal{R}^{\rm N}_{\odot}$}}

\newcommand{\Lnom}{\hbox{$\mathcal{L}^{\rm N}_{\odot}$}}
\newcommand{\Tnom}{\hbox{$\mathcal{T}^{\rm N}_{\rm eff\odot}$}}
\newcommand{\GMnom}{\hbox{$\mathcal{(GM)}^{\rm N}_{\odot}$}}
\newcommand{\GMJnom}{\hbox{$\mathcal{(GM)}^{\rm N}_{\rm J}$}}
\newcommand{\GMEnom}{\hbox{$\mathcal{(GM)}^{\rm N}_{\rm E}$}}
\newcommand{\ReJnom}{\hbox{$\mathcal{R}^{\rm N}_{e\rm J}$}}
\newcommand{\RpJnom}{\hbox{$\mathcal{R}^{\rm N}_{p\rm J}$}}
\newcommand{\ReEnom}{\hbox{$\mathcal{R}^{\rm N}_{e\rm E}$}}
\newcommand{\RpEnom}{\hbox{$\mathcal{R}^{\rm N}_{p\rm E}$}}

\newcommand{\ls}{$L_{\odot}$}
\newcommand{\rs}{$R_{\odot}$}

\newcommand{\MnomJ}{\hbox{$\mathcal{M}_{\rm J}^{\rm N}$}}
\newcommand{\MnomE}{\hbox{$\mathcal{M}_{\rm E}^{\rm N}$}}

\begin{document}

\title{Resolution B3\\
  on recommended nominal conversion constants for
  selected solar and planetary properties\\
{\small\it Proposed by IAU Inter-Division A-G Working Group on Nominal
  Units for Stellar \& Planetary Astronomy}}
\maketitle


\noindent The XXIXth International Astronomical Union General Assembly,

\bigskip
\noindent {\bf Recognizing}

\bigskip

\noindent that notably different values of the solar mass, radius,
luminosity, effective temperature, total solar irradiance, of the
masses and radii of the Earth and Jupiter, and of the Newtonian
constant of gravitation $G$ have been used by researchers to express
and derive fundamental stellar and planetary properties,


\bigskip

\noindent {\bf Noting}

\bigskip

\begin{enumerate}

\item that neither the solar nor the planetary masses and radii are
  secularly constant and that their instantaneous values are gradually
  being determined more precisely through improved observational
  techniques and methods of data analysis, and

\item that the common practice of expressing the stellar and planetary
  properties in units of the properties of the Sun, the Earth, or
  Jupiter inevitably leads to unnecessary systematic differences that
  are becoming apparent with the rapidly increasing accuracy of
  spectroscopic, photometric, and interferometric observations of
  stars and extrasolar planets\endnote{~Note, e.g., that since
    projected rotational velocities of stars ($v\,\sin\,i$) are
    measured in SI units, the use of different values for the solar
    radius can lead to measurable differences in the rotational
    periods of giant stars (see Harmanec and Pr\v{s}a 2011).}, and
   
\item that the universal constant of gravitation $G$ is currently one
  of the least precisely determined constants, whereas the error in
  the product $GM_{\odot}$ is five orders of magnitude smaller (Petit
  \& Luzum 2010, and references therein),
  
\end{enumerate}
  

\bigskip

\noindent {\bf Recommends}

\bigskip

\noindent In all scientific publications in which {\bf accurate}
values of basic stellar or
planetary properties are derived or quoted:

\begin{enumerate}

\item that whenever expressing stellar properties in units of the
  solar radius, total solar irradiance, solar luminosity, solar
  effective temperature, or solar mass parameter, that the nominal
  values \Rnom, \Snom, \Lnom, \Tnom, and \GMnom, be used,
  respectively, which are by definition {\it exact} and are expressed
  in SI units.  These {\it nominal} values should be understood as
  conversion factors only --- chosen to be close to the current
  commonly accepted estimates (see table below) --- not as the true
  solar properties.  Their consistent use in all relevant formulas
  and/or model calculations will guarantee a uniform conversion to SI
  units. Symbols such as \ls\ and \rs, for example, should only be
  used to refer to actual estimates of the solar luminosity and solar
  radius (with uncertainties),

\item that the same be done for expressing planetary properties in
  units of the equatorial and polar radii of the Earth and Jupiter
  (i.e., adopting nominal values \ReEnom, \RpEnom, \ReJnom, and
  \RpJnom, expressed in meters), and the nominal terrestrial and
  jovian mass parameters \GMEnom\ and \GMJnom, respectively (expressed
  in units of m$^{3}$\,s$^{-2}$). Symbols such as GM$_{\rm E}$, listed
  in the IAU 2009 system of astronomical constants (Luzum et
  al. 2011), should be used only to refer to actual estimates (with
  uncertainties),

\item that the IAU (2015) System of Nominal Solar and Planetary
  Conversion Constants be adopted as listed below:

\begin{center}
\begin{tabular}{p{2cm}p{1cm}p{8cm}}
\hline
\multicolumn{3}{c}{SOLAR CONVERSION CONSTANTS} \\
\hline
$1 \Rnom$ & = & $6.957 \times 10^8$\,m \\
$1 \Snom$ & = & $1361$\,W\,m$^{-2}$\\
$1 \Lnom$ & = & $3.828 \times 10^{26}$\,W \\
$1 \Tnom$ & = & $5772$\,K\\
$1 \GMnom$& = & $1.327\,124\,4 \times 10^{20}\ \mathrm{m}^3 \mathrm{s}^{-2}$\\
\hline
\end{tabular}

\smallskip

\begin{tabular}{p{2cm}p{1cm}p{8cm}}
\hline
\multicolumn{3}{c}{PLANETARY CONVERSION CONSTANTS} \\
\hline
$1\,\ReEnom$ & = & $6.3781\,\times\,10^{6}$\,m \\
$1\,\RpEnom$ & = & $6.3568\,\times\,10^{6}$\,m \\
$1\,\ReJnom$ & = & $7.1492\,\times\,10^{7}$\,m \\
$1\,\RpJnom$ & = & $6.6854\,\times\,10^{7}$\,m \\
$1\,\GMEnom$ & = & $3.986\,004\,\times\,10^{14}\,\mathrm{m}^{3}\,\mathrm{s}^{-2}$\\
$1\,\GMJnom$ & = & $1.266\,865\,3\,\times\,10^{17}\,\mathrm{m}^{3}\,\mathrm{s}^{-2}$\\
\hline
\end{tabular}
\end{center}


\item that an object's mass can be quoted in nominal solar masses
  ${\mathcal M^{\rm N}_{\odot}}$ by taking the ratio $(GM)_{\rm
  object}$/\GMnom, or in corresponding nominal jovian and terrestrial
  masses, \MnomJ\, and \MnomE, respectively, dividing by \GMJnom\, and
  \GMEnom,

\item that if SI masses are explicitly needed, they should be
  expressed in terms of $(GM)_{\rm object}/G$, where the estimate of
  the Newtonian constant $G$ should be specified in the publication
  (for example, the 2014 CODATA value is $G = 6.67408\,(\pm0.00031)
  \times 10^{-11}$\,m$^3$\,kg$^{-1}$\,s$^{-2}$),

\item that if nominal volumes are needed, that a nominal terrestrial
  volume be derived as $4\,\pi\,\ReEnom^{2}\,\RpEnom/3$, and nominal
  jovian volume as $4\,\pi\,\ReJnom^{2}\,\RpJnom/3$.
  
\end{enumerate}

\bigskip

\noindent {\bf Explanation}

\bigskip

\begin{enumerate}
  
\item The need for increased accuracy has led to a requirement to
  distinguish between Barycentric Coordinate Time (TCB) and
  Barycentric Dynamical Time (TDB).  For this reason the {\it nominal
    solar mass parameter} \GMnom\ value is adopted as an exact number,
  given with a precision within which its TCB and TDB values agree
  (Luzum et al. 2011).  This precision is considered to be sufficient
  for most applications in stellar and exoplanetary research for the
  forseeable future.

\item The {\it nominal solar radius} \Rnom\ corresponds to the solar
  photospheric radius suggested by Haberreiter et
  al.\ (2008)\endnote{~Haberreiter et al.\ (2008) determined the solar
    photospheric radius, defined to be where $\tau_{\rm Ross}$ = 2/3,
    to be 695\,658\,($\pm$\,140)\,km. The adopted
    \Rnom\ is based on this value, quoting an appropriate number of
    significant figures given the uncertainty, and differs slightly
    from the nominal solar radius tentatively proposed by Harmanec \&
    Pr\v{s}a (2011) and Pr\v{s}a \& Harmanec (2012).}, who resolved
  the long-standing discrepancy between the seismic and photospheric
  solar radii.  This \Rnom\ value is consistent with that adopted by
  Torres et al.\ (2010) in their recent compilation of updated radii
  of well observed eclipsing binary systems.

\item The {\it nominal total solar irradiance} \Snom\ corresponds to
  the mean total electromagnetic energy from the Sun, integrated over
  all wavelengths, incident per unit area per unit time at distance
  1\,au --- also measured contemporarily as the {\it total solar
    irradiance} (TSI; e.g., Willson 1978) and known historically as
  the {\it solar constant} (Pouillet 1838). \Snom\ corresponds to the
  solar cycle 23-averaged TSI
  ($S_{\odot}$\,=\,1361\,($\pm$\,1)~W\,m$^{-2}$; 2$\sigma$
  uncertainty; Kopp et al., in prep.)\endnote{~The TSI is variable at
    the $\sim$0.08\%\, ($\sim$1 W\,m$^{-2}$) level and may be variable
    at slightly larger amplitudes over timescales of centuries. Modern
    spaceborne TSI instruments are absolutely calibrated at the
    0.03\%\, level (Kopp 2014). The TIM/SORCE experiment established a
    lower TSI value than previously reported based on the fully
    characterized TIM instrument (Kopp et al. 2005, Kopp \& Lean
    2011). This revised TSI scale was later confirmed by
    PREMOS/PICARD, the first spaceborne TSI radiometer that was
    irradiance-calibrated in vacuum at the TSI Radiometer Facility
    (TRF) with SI-traceability prior to launch (Schmutz et al. 2013).
    The SOVAP/PICARD (Meftah et al. 2014), ACRIM3/ACRIMSat (Willson
    2014), VIRGO/SoHO, and TCTE/STP-Sat3
    (http://lasp.colorado.edu/home/tcte/) flight instruments are now
    consistent with this new TSI scale within instrument
    uncertainties, with the DIARAD, ACRIM3, and VIRGO having made
    post-launch corrections and the TCTE having been validated on the
    TRF prior to its 2013 launch. The Cycle 23 observations with these
    experiments are consistent with a mean TSI value of
    $S_{\odot}$\,=\,1361\,~W\,m$^{-2}$\,($\pm$\,1~W\,m$^{-2}$;\,2$\sigma$).
    The uncertainty range includes contributions from the absolute
    accuracies of the latest TSI instruments as well as uncertainties
    in assessing a secular trend in TSI over solar cycle 23 using
    older measurements.}.

\item The {\it nominal solar luminosity} \Lnom\ corresponds to the
  mean solar radiative luminosity rounded to an appropriate number of
  significant figures. The current (2015) best estimate of the mean
  solar luminosity $L_{\odot}$ was calculated using the solar
  cycle-averaged TSI$^3$ and the IAU 2012 definition of the
  astronomical unit\endnote{~Resolution B2 of the XXVIII General
    Assembly of the IAU in 2012 defined the astronomical unit {\it to
      be a conventional unit of length equal to 149\,597\,870\,700\,m
      exactly}. Using the current best estimate of the TSI (discussed
    in endnote 3), this is consistent with a current best estimate of
    the Sun's mean radiative luminosity of $L_{\odot}$\,=\,
    4\,$\pi$\,(1\,au)$^2$\,$S_{\odot}$\,
    =\,3.8275\,($\pm$\,0.0014)\,$\times\,10^{26}$\,W.}.

\item The {\it nominal solar effective temperature} \Tnom\ corresponds
  to the effective temperature calculated using the current (2015)
  best estimates of the solar radiative luminosity and photospheric
  radius, and the CODATA 2014 value for the Stefan-Boltzmann
  constant\endnote{~The CODATA 2014 value for the Stefan-Boltzmann
    constant is $\sigma = 5.670\,367\,(\pm\,0.000\,013) \times
    10^{-8}$\,W\,m$^{-2}$\,K$^{-4}$.  The current best estimate for
    the solar effective temperature is calculated to be $T_{\rm
      eff,\odot}$\,=\,5772.0\,($\pm$\,0.8)\,K.}, rounded to an
  appropriate number of significant figures.

\item The parameters \ReEnom\, and \RpEnom\, correspond respectively
  to the Earth's ``zero tide" equatorial and polar radii as adopted
  following 2003 and 2010 IERS Conventions (McCarthy \& Petit 2004;
  Petit \& Luzum 2010), the IAU 2009 system of astronomical constants
  (Luzum et al.\ 2011), and the IAU Working Group on Cartographic
  Coordinates and Rotational Elements (Archinal et al.\ 2011). If
  equatorial vs. polar radius is not explicitly specified, it should
  be understood that {\it nominal terrestrial radius} refers
  specifically to \ReEnom, following common usage.

\item The parameters {\ReJnom} and \RpJnom\ ~correspond respectively
  to the one-bar equatorial and polar radii of Jupiter adopted by the
  IAU Working Group on Cartographic Coordinates and Rotational
  Elements 2009 (Archinal et al. 2011). If equatorial vs. polar radius
  is not explicitly specified, it should be understood that {\it
    nominal jovian radius} refers specifically to \ReJnom, following
  common usage.

\item The {\it nominal terrestrial mass parameter} {\GMEnom}\, is
  adopted from the IAU 2009 system of astronomical constants (Luzum et
  al. 2011), but rounded to the precision within which its TCB and TDB
  values agree. The {\it nominal jovian mass parameter} \GMJnom\, is
  calculated based on the mass parameter for the Jupiter system from
  the IAU 2009 system of astronomical constants (Luzum et al. 2011),
  subtracting off the contribution from the Galilean satellites
  (Jacobson et al. 2000).  The quoted value is rounded to the
  precision within which the TCB and TDB values agree, and the
  uncertainties in the masses of the satellites are negligible.

\item The nominal value of a quantity $Q$ can be transcribed in LaTeX
  with the help of the definitions listed below for use in the text
  and in equations:

\smallskip
\begin{verbatim}
\newcommand{\Qnom}{\hbox{$\mathcal{Q}^{\rm N}_{\odot}$}}
\newcommand{\Qn}{\mathcal Q^{\rm N}_{\odot}}
\end{verbatim}
  
\end{enumerate}

\bigskip

\noindent {\bf References}

\smallskip

\noindent Archinal, B.\ A., A'Hearn, M.\ F., Bowell, E., et al.\ 2011,
Celestial Mechanics and Dynamical Astronomy 109, 101

\smallskip

\noindent Haberreiter, M., Schmutz, W., Kosovichev, A.\ G. 2008, ApJ,
675, L53

\smallskip

\noindent Harmanec, P., Pr\v{s}a, A. 2011, PASP, 123, 976

\smallskip

\noindent Jacobson, R.~A., Haw, R.~J., McElrath, T.~P., \&
Antreasian, P.~G. 2000, J. Astronaut. Sci. 48(4), 495

\smallskip

\noindent Kopp, G. 2014, Journal of Space Weather and Space Climate,
4, A14

\smallskip

\noindent Kopp, G., Lawrence, G., Rottman, G., 2005, Solar Physics,
230, 129

\smallskip

\noindent Kopp, G., \& Lean, J.~L. 2011, Geophys. Res. Letters, 38,
L01706

\smallskip

\noindent Luzum, B., Capitaine, N., Fienga, A., et al.\ 2011,
Celestial Mechanics and Dynamical Astronomy, 110, 293

\smallskip

\noindent McCarthy, D.\ D. \& Petit, G. 2004 IERS Technical Note
No. 32, 1

\smallskip

\noindent Meftah, M., Dewitte, S., Irbah, A., et al.\ 2014, Solar
Physics, 289, 1885

\smallskip

\noindent Petit, G., Luzum, B.\ (Eds.) 2010 IERS Technical Note
No. 36

\smallskip

\noindent Pouillet, C.~S.~M.\ 1838, {\it Memoire sur le chaleur
  solaire}, Paris, Bachelier

\smallskip

\noindent Pr\v{s}a, A. \& Harmanec, P. 2012, Proc.\ IAU Symp.\ 282,
Cambridge Univ., Press, 339

\smallskip

\noindent Schmutz W., Fehlmann A., Finsterle W., et al.\ 2013,
AIP Conf. Proc. 1531, p. 624–627, doi:10.1063/1.4804847

\smallskip

\noindent Torres, G., Andersen, J., Gim\'enez, A. 2010, A\&A Rev., 18,
67

\smallskip

\noindent Willson, R.\ C. 2014, Astrophysics \& Space Science, 352,
341

\smallskip

\noindent Willson, R.~C.\ 1978, Journal of Geophysical Research, 83,
4003

\theendnotes

\end{document}